# Raman study on G mode of graphene for determination of edge orientation

*Chunxiao Cong, Ting Yu\*, and Haomin Wang*

*Division of Physics and Applied Physics, School of Physical and Mathematical Sciences, Nanyang Technological University, Singapore 637371, Singapore*

*Address correspondence to yuting@ntu.edu.sg

**ABSTRACT**

We report a confocal Raman study on edges of single layer graphene. It is found that edge orientations could be identified by G mode besides D mode. We observe that G mode at edges of single layer graphene exhibits polar behaviors and different edges like zigzag- or armchair-dominated responses differently to the polarization of the incident laser. Moreover, G mode shows stiffening at zigzag-dominated edges, while it is softened at armchair- dominated ones. Our observations are in good agreement with recent theory (K. Sasaki, *et al*., J. Phys. Soc. Jpn. 79, 044603) and could be well explained by the unique properties of pseudospin at graphene edges, which lead to asymmetry of Raman active modes and non-adiabatic processes (Kohn Anomaly) at different types of edges. This work could be useful for further study on the properties of graphene edge and development of graphene-based devices.

**KEYWORDS:** graphene   edge orientation   Raman   polarization

Graphene, the mother material of other sp2 carbon allotropes became the most shining star in carbon family since it was discovered experimentally and its quantum Hall properties were reported. [1-3] Among the research interests of graphene, edge is one of the focuses because physical properties of graphene, especially graphene nanoribbons are mainly determined by edges. Some edges with certain symmetry may have localized edge states while others do not. [4, 5] The superconductivity, [6] ferromagnetism [4] and exceptional quantum Hall effect [7] might be resulted from these localized edge states. Different types of edges may also have various favorites to functional chemical and biological groups, which could certainly extend the potentials of graphene. [8]

Raman spectroscopy is a unique and one of the most powerful techniques for graphene studies. Usually, two strong peaks are presented in Raman spectrum of graphene of large dimension and high quality, named G and G' (or 2D) modes. If laser beam is focused on edges of graphene or some defects exist, another peak will appear, called D mode. All Raman modes of graphene are resonant, meaning phonons represented by these modes are strongly interacted with electron and hole through the famous single resonance for G mode and intervalley double resonance for D and G' modes. [9] An elastic scattering by defects is required for observation of D mode. [10] Thus, by monitoring the position, width, integrated intensity and so on, the number of layers, [11, 12] linear dispersion of electronic energy, [13] crystal orientation, [14, 15] doping, [16, 17] defects [18] and strain [19, 20] could be probed by Raman study. Most recently, we demonstrated that the armchair-dominated edge could be distinguished from the zigzag-dominated edge by reading the integrated intensity of D mode. This could be well explained by the defect assisted double resonant intervalley scattering process. [21, 22]

In this paper, we demonstrate the orientation identification of graphene edge through G mode. In contrast to the body of graphene sheet, the G mode at edge exhibits obvious polarization dependence. The opposite polar behaviors of armchair edge (*A*-edge) and zigzag edge (*Z*-edge) could be an alternative way to effectively identifying the orientation of graphene edges. In addition, the presence and removal of non-adiabatic Kohn anomaly have also been seen at *A*- and *Z*- edge, respectively, which is also an option for determining edge orientation.

**Results and Discussion**

Considering the crystal nature of graphene, only SLG sheets having edges with a relative angle of multiple of 30° were selected. As shown in Fig. 1A, a 90° corner between two adjacent edges can be clearly seen in the optical image. Two peaks, named G and G' modes are observed when the laser beam was focused on the body of the SLG sheet (inset of figure S1A). By extracting the integrated intensity of G mode, a Raman image of the interested corner is constructed (Fig. 1B) and exhibits a significant contrast between the main body and edges, indicating a dramatic decrease of G mode intensity moving from inside to outside. The profile of the G mode intensity across the edges reveals step-function dependence as shown in Figure S1A, which agrees well with previous report.[23] Differing from G mode, D mode originating from a defect assisted double resonance process localizes at one edge (Fig. 1C) when the polarization of the incident laser is tuned to be 45° with respect to each of edges. The localization of D mode could be well demonstrated by the Gaussian profile (Fig. S1B). The strong D mode at the vertical edge and very weak D mode at the horizontal edge suggest the *A*- and *Z*-edge nature of the edges, respectively, because *A*-edge could scatter electrons from one to the other valley near the Dirac points – so called intervalley scattering - whereas *Z*-edge cannot.[22] Hereinafter, we label the vertical edge as armchair and the horizontal as zigzag.

Considering the fact that width of the edge for G mode is very narrow as the consequence of zero momentum of such G phonons,[24] and to effectively collect enough signal from the edge, polarization dependent Raman mapping of G mode is conducted. As shown in Figure 2A-bottom, with increasing the angles of the incident electric field relative to the *A*-edge, the G modes become weaker and minimum at 90°, which is consist with theoretical results.[25] As comparison, the polar behavior of D mode is also revealed by a group of polar-Raman images shown in Fig. 2A-top. It can be clearly seen that the intensity of D mode decreases with the increase of the angles. As known, the intensity of D mode is proportional to [**e** X **p**], here **e** is the polarization of the incident laser, and **p** is the momentum of photoexcited electron or hole relative to the Dirac point. As such, the strongest D mode appears when the polarization of the incident light is parallel to the A-edge. Essentially, the intensity of D mode follows $I_D \propto \cos^2(\theta_{in})\cos^2(\theta_{out})$, here $\theta_{in}$ and $\theta_{out}$ represent the angles of the incident and scattering electric fields relative to the edge, respectively. [23, 24] In this work, we did not put an analyzer in the scattering beam. Therefore, the intensity of D mode is only proportional to $\cos^2(\theta_{in})$, as shown in Figure S2. A ratio of 0.2 between the $I_{Dmin}$ to $I_{Dmax}$ is obtained, implying a high purity and degree of order of this *A*-edge.[24] Figure 2B shows the typical Raman spectra of G mode at the *A*-edge with various polar angles. The inset plots the corresponding Raman spectra taken from the body. This polarization independence is directly resulted from the nature of the G mode in graphene, which has an $E_{2g}$ symmetry (space group $P6_3/mmc$) [26] and the two doubly degenerate Raman polarizability tensors functioning like reflecting the polarization vector about the x = y and y = 0 with identical scattering efficiency. [14, 23, 27]

Figure 3 shows the polarization dependence of both G and D modes at the Z-edge. Obviously, the polar behavior of the G mode at the Z-edge differs strongly from that at the *A*-edge. The larger polar angles lead to stronger G mode in Z-edge as predicted by the theoretical work. [25] Weak D mode caused by the existence of some armchair segments in the Z-edge still obeys the standard dependence. The typical Raman spectra of G mode at the zigzag edge are shown in Figure 3B.

To further elucidate the polarization dependence of G mode at both *A*- and Z-edge, the intensities of G mode are plotted as a function of polar angles as shown in Figure 4. Since the G mode at the main body has no polarization dependence, the data were normalized to the $I_G$ at the body. For *A*-edge (Figure 4A), $\cos^2(\theta_{in})$ is employed to fit the experimental data whereas the intensity of G mode at Z-edge (Figure 4B) follows $\sin^2(\theta_{in})$ perfectly. The appearance of remarkable polar behavior at the edges differs sharply from that at the body. This is because the two degenerate components of G mode, longitudinal optical (LO) and transverse optical (TO) phonons no longer participate the scattering process equally. In fact, only LO mode is active at armchair edge while TO mode is dominated at zigzag edge.[25]

Figure 5A is an optical image of a SLG sheet with a 90° (labeled as 1), a 120° (labeled as 2) and a 330° (labeled as 3) corner. Figure 5B-left shows the Raman mapping of D mode of these three corners. As expected, at 90° and 330° corners, one edge presents stronger D mode while the other exhibits weaker one, whereas, the two edges with an angle of 120° show almost the same contrast. Based on the previous analysis, the edge orientations are identified as illustrated in Fig. 5B-left. Interestingly, when we plotted the images of G mode to show the distribution of G peak position over the corners (Fig. 5B-middle), we noticed that the zigzag edges appear brighter comparing to the other part, meaning a blue-shift occurs to the G mode at the zigzag edges. Two mechanisms could be responsible for such blue-shift for the zigzag edge or relatively red-shift for the armchair edge and main body. One is uniaxial strain and the other is Kohn anomaly. Tensile (compressive) strain is able to lower (increase) the frequency of G phonons (both LO and TO components).[28] In addition to G mode, strain could also cause the shift of G' mode and actually G' mode is more sensitive to the strain.[20] So, we examined the G' mode by the Raman mapping of G' mode position (Figure 5B-right). Among three images, only the 330° corner shows an obvious blue-shift at the Z-edge which is coincident with the G mode and implies the strain might be one of the reasons resulting in such blue-shift happened to this edge. Since the images of the 90° and 120° corners present quite uniform distribution of the position of G' mode over the entire areas, we should consider the other possible process causing such difference of G peak position -Kohn anomaly- a process capable of softening Γ point phonons (G mode in graphene) with a zero momentum through a phonon electron interaction.[29] Due to the unique crystal configurations at A- and Z-edge, the TO phonons at both zigzag and armchair almost have no contribution to the creation of electron-hole, a critical step in the Kohn anomaly, which means that TO mode cannot undergo the Kohn anomaly as contrast to the LO mode.[29] Previous analyses have revealed that LO mode is only active at A-edge. Hence, the Kohn anomaly is expected to happen to G mode at the A-edge and lead to a red-shift of G mode as what we observed (Fig. 5B-middle). The relative blue-shift of G mode at Z-edge is the consequence of breaking Kohn anomaly since the LO mode is not active at the Z-edge.

In summary, we studied the G mode at the edges of SLG sheets by Raman spectroscopy and mapping. Different polarization dependence of G mode at *Z*- and *A*-edge (pre-determined by the D mode) has been observed and could be well explained by the unequally interaction between LO and TO phonons with electrons at the different types of edges. Kohn anomaly at the *A*-edge and breaking of Kohn anomaly at the *Z*-edge have been demonstrate by the Raman mapping of G mode position. Further studies on this Kohn anomaly at the edge of graphene are needed as we did observe the coincidence of blue-shift of both G and G' mode, a likely result of strain effect.

**Experimental Methods**

Graphene samples were prepared by micromechanical cleavage of natural graphite crystals and transferred on Si wafer with 300 nm oxide cap layer. Single layer graphene (SLG) flakes were selected by reading the width of G' mode.[11] Raman spectra and images were obtained by a WITeck CRM200 confocal microscopy Raman system with a piezocrystal controlled scanning stage. The excitation light is 532 nm laser. The laser spot size is estimated to be 500 nm. To avoid damaging and heating, the laser was attenuated and the power is controlled below 2mW. The fast scan rate 1 sec per pixel also helped to minimize the laser effects. The polarization of the incident laser was tuned by rotating a half-wave plate.

**Supporting Information Available**

10 pieces of samples have been examined in total. 7 samples show clearly contrast of D mode intensity between two adjacent edges and 3 samples among these 7 present different polarization dependence of G mode and/or Kohn anomaly. The relatively low frequency of observing G mode polar behavior and Kohn anomaly could be due to the fact that edges of graphene are more sensitive to other factors such as strain and doping which could effectively change the G mode position and intensity. Furthermore, the effective edge for G mode is very narrow and the G mode from the body is much stronger than that from the edge which makes it extremely hard to collect signal purely from the edges. This is also one of our motivations to employ Raman mapping which is able to provide more statistical data.

FIGURE CAPTIONS

**Figure 1** (A) Optical image of SLG with two edges at 90°. (B) and (C) are Raman images of the interested area highlighted in (A) by extracting the integrated intensity of G and D modes, respectively. The polarization of the incident laser is at 45° to each of edges as illustrated.

**Figure 2** Raman images of D (A-top) and G (A-bottom) intensities. The polarization of the incident laser is tuned to be 0°, 30°, 60° and 90° with respect to the armchair edge. (The scale bar is 400 nm.) (B) Typical Raman spectra of the armchair edge showing the polarization dependence of G mode. The inset plots the corresponding Raman spectra taken from the body.

**Figure 3** Raman images of D (A-left) and G (A-right) intensities. The polarization of the incident laser is tuned to be 0°, 30°, 60° and 90° with respect to the average direction of the zigzag edge. (The scale bar is 400 nm.) (B) Typical Raman spectra of the zigzag edge showing the polarization dependence of G mode.

**Figure 4** Polar plots of the integrated intensity of G mode at the (A) *A*-edge, and (B) *Z*- edge. The red solid line in (A) and (B) is the fitting curve of $I_G \propto \cos^2(\theta_{in})$ and $I_G \propto \sin^2(\theta_{in})$ of the experimental data, respectively.

**Figure 5** (A) Optical image of SLG with a 90º, a 120º and a 330º corner as labeled by (1), (2) and (3), respectively. (B) Raman images of integrated intensity of D mode (left), position of G (middle) and position of G' modes (right), respectively.

**Figure S1** Integrated intensity of G mode (A) and D mode (B) as a function of position across the edge from inside to outside. The insets of (A) and (B) are the typical Raman spectrum taken from the body and the edge of SLG, respectively.

**Figure S2** Polar plots of the integrated intensity of D mode at the *A*-edge. The red solid line is the fit of experimental data to curve of $\sin^2(\theta_{in})$.

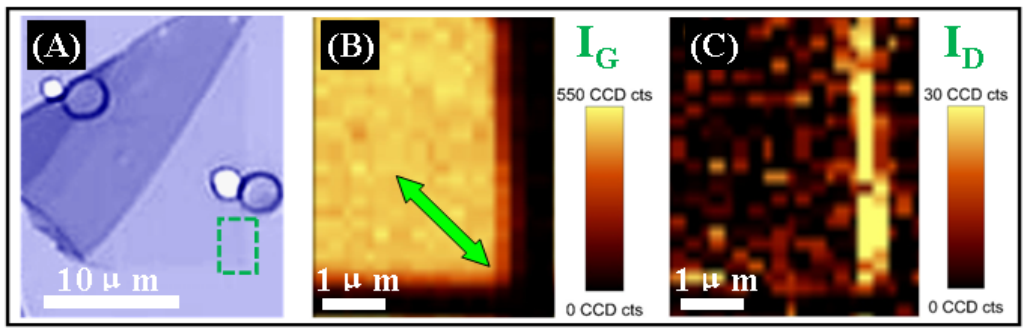

**Figure 1**

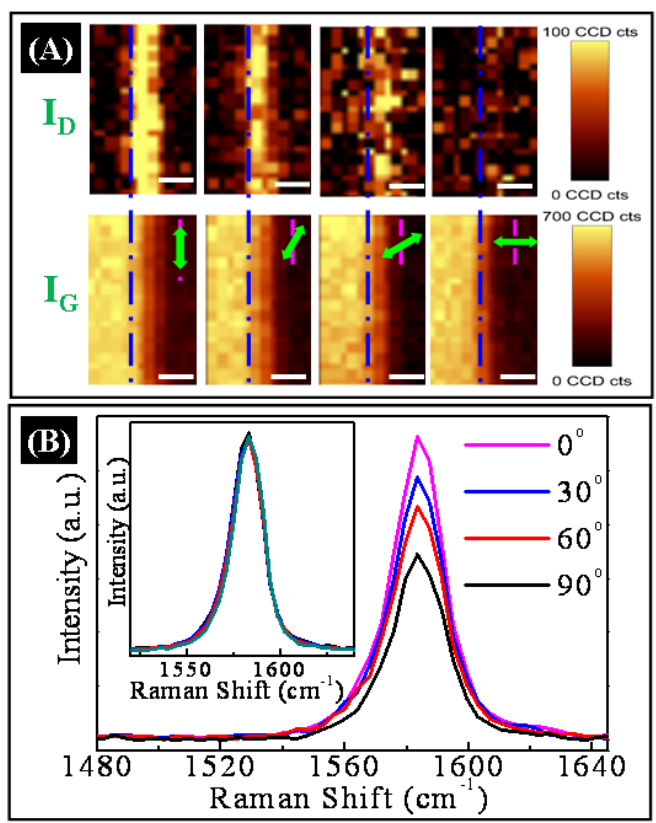

Figure 2

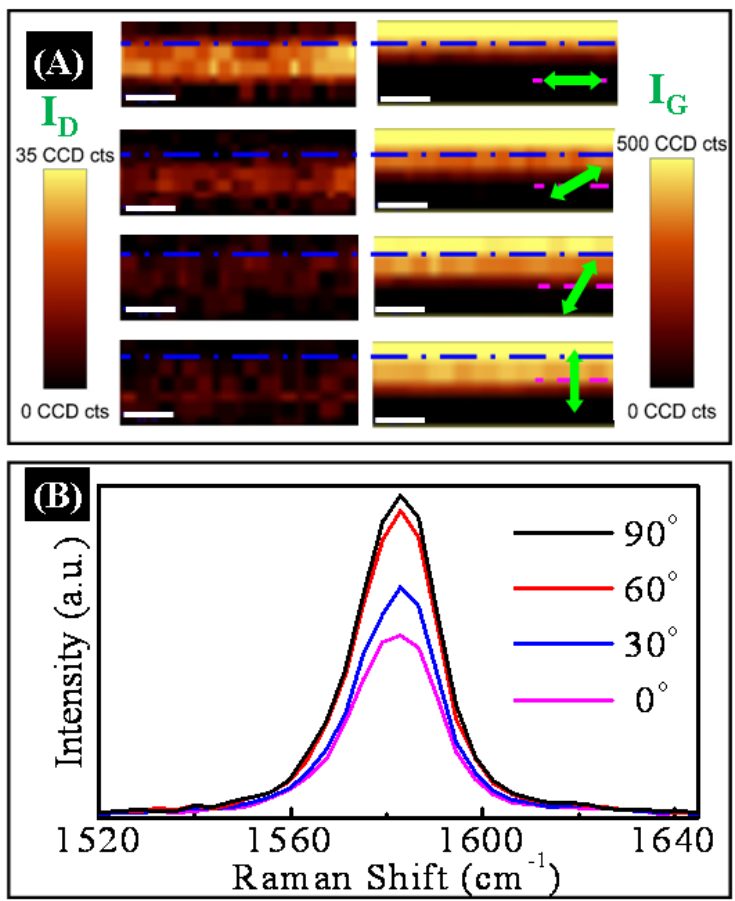

**Figure 3**

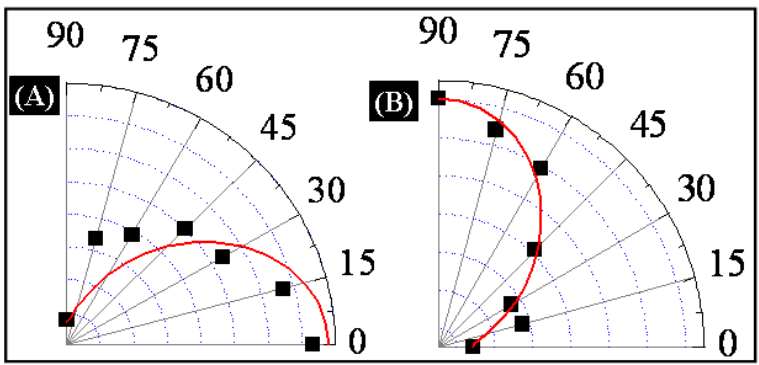

**Figure 4**

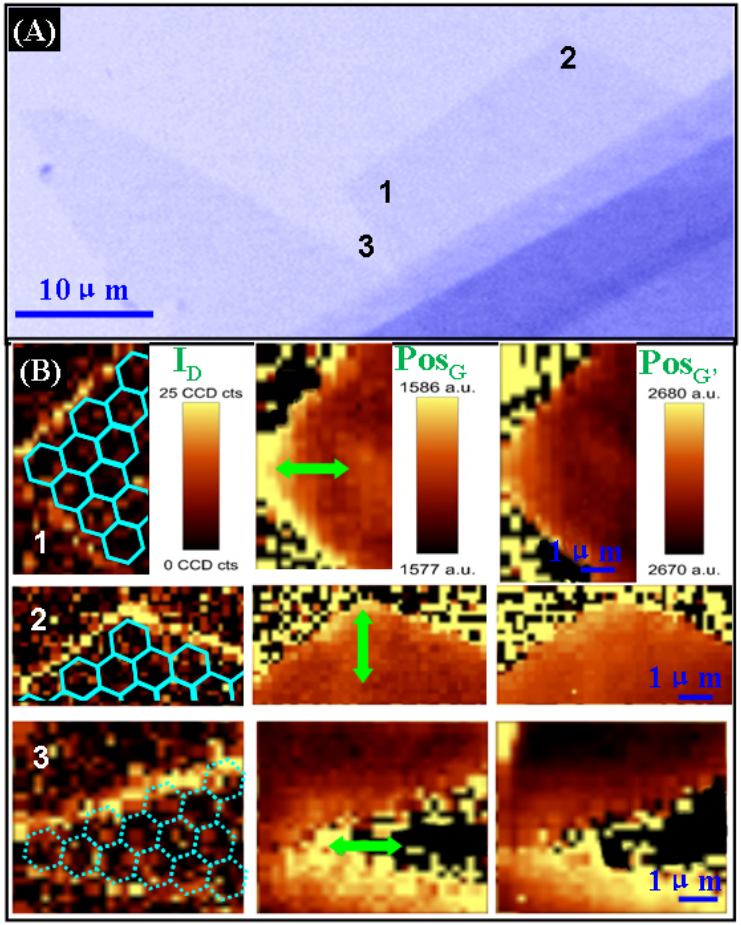

**Figure 5**

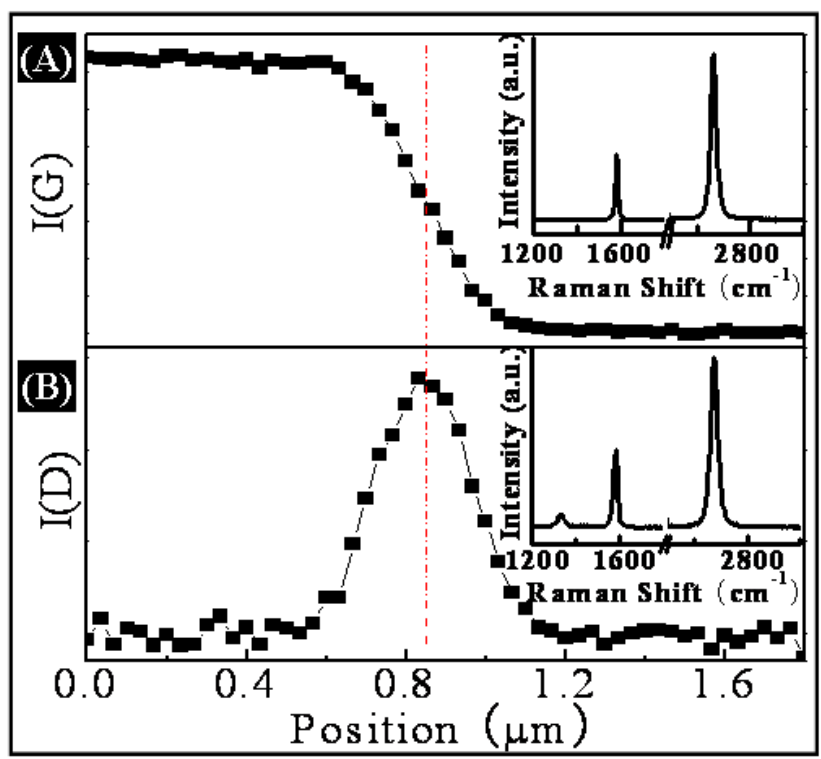

**Figure S1**

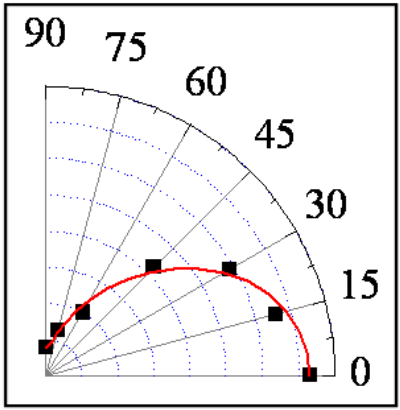

**Figure S2**